# X-ray view on a Class using Conceptual Analysis in Java Environment

Gulshan Kumar [1], Mritunjay Kumar Rai[2]

[1]Department of Computer Science and Engineering, Lovely Professional University

Phagwara, Punjab, India

gulshan_acet@yahoo.com

[2]Department of Computer Science and Engineering, Lovely Professional University

Phagwara, Punjab, India[2]

raimritunjay@gmail.com

**Abstract**

Modularity is one of the most important principles in software engineering and a necessity for every practical software. Since the design space of software is generally quite large, it is valuable to provide automatic means to help modularizing it. An automatic technique for software modularization is object- oriented concept analysis (OOCA). X-ray view of the class is one of the aspect of this Object oriented concept analysis. We shall use this concept in a java environment.

## 1. Introduction

Concept Analysis (CA) is a branch of lattice theory that allows us to identify meaningful groupings of elements (referred to as objects in CA literature) that have common properties (referred to as attributes in CA literature). These groupings are called concepts and capture similarities among a set of elements based on their common properties. In the specific case of software reengineering, the system are composed of a big amount of different entities (classes, methods, modules, subsystems) and there are different kinds of relationships among them. It also represents dependencies among the classes or entities.

X-Ray views —a technique based on Concept Analysis— which reveal the internal relationships between groups of methods and attributes of a class. X-Ray views are composed out of elementary collaborations between attributes and methods and help the engineer to build a mental model of how a class works internally.

## 2. Existing Idea

Within object oriented software, the minimal unit of development and testing is a class. Usually, a class is composed of instance variables used to represent the state, and methods used to represent the behavior of the classes. Then, understanding how a class works means identifying several aspects:

How the methods are interacting together (coupling between methods)

How the instance variables are working (or not) together in the methods (coupling between instance variables)

Which methods are using (or not) the state of the class

if there are methods that form a cluster and define together a precise behaviour of the class

Which methods are considered as interfaces

Which methods are used as entry points (methods that are considered as interfaces and communicate with other methods defined in the class)

Which methods and instance variables represent the core of the class

Which methods are using all the state of the class

In paper [1], the authors have given an idea of concept analysis. Mathematically, concepts are maximal collections of elements sharing common properties. To use the CA technique, one only needs to specify the properties of interest on each element, and does not need to think about all possible combination of these properties, since these groupings are made automatically by the CA algorithm. The possibility of capturing similarities of elements in groups (concepts) -based on the specification of simple properties allow to identify common features of the elements. When we are able to characterize the entities in terms of properties, and we can detect if these characteristics are repeated in the system, then we can reduce the amount of information to analyze and we can have an abstraction of the different parts of a system. These abstractions help us to start to see how the parts are working, how they are defined and how they are connected to other parts of the system .The elements are the instance variables and the methods defined in the class, and the properties are how they are related between themselves.

If we have the set of instance variables {A, B}, and the set of methods {P, Q, X, Y} defined in a class, the properties we use are:
B is used by P means that the method P is accessing directly or through an accessor / mutator to the instance variable B.
Q is called in P means that the method Q is called in the method P via a self-call. It also shows indirect dependencies between elements if exists. They have also shown different types of relations and dependencies through some notations.

{E1, ..,En} R {M1, ..,Mp} means that the entities {E1, ..,En} depend exclusively on {M1, ..,Mp}. This means that {M1, ..,Mp} are the only entities that are related through the property R to {E1, ..,En}.

{E1, ..,En} R {M1, ..,Mp} means that the entities {E1, ..,En} do not depend exclusively on {M1, ..,Mp}.

{E1, ..,En} $\overline{R}$*{M1, ..,Mp} means that the entities {E1, ..,En} depend exclusively and transitively on {M1, ..,Mp}. This means that {M1, ..,Mp} are the only ones that are related to {E1, ..,En} through the property R and R1, where R1 is an intermediate property, because there is a set {N1, ..,Nk} such that: {E1,..,En} R {N1, ..,Nk} R1 {M1, ..,Mp}.

{E1, ..,En} $R^{\perp}$ {M1, ..,Mp} means that the entities {E1, ..,En} do not depend exclusively but transitively on {M1, ..,Mp}. This means that {M1, ..,Mp} are not the only ones that are related to {E1, ..,En} through the property R and R1, where R1 is an intermediate property, because there is a set {N1, ..,Nk} such that: {E1, ..,En} R {N1, ..,Nk} R1 {M1, ..,Mp}.

A special case: {E1, ..,En} ¬R {M1, ..,Mp} means that the entity {E1, ..,En} has any dependencies on {M1, ..,Mp}. This is only applicable on exclusive dependencies.

In paper [2], the authors have discussed different types of X-ray views which will be helpful for our future work. In paper [3] there is a concept on modularization using the conceptual analysis on object oriented environment.

## 3. Our Application

Our idea is now to use the above said concepts in the environment of java and to way out the modularization in java programs. Modularization also helps in software re-engineering.
For the present purpose, let us have an example of java coding. We have applied the proposed idea in different properties of Java programming each of which are illustrated below.

3.1 Polymorphism

Polymorphism deals with of different forms of a method where parameters are different according to the forms of the methods. Polymorphism can also occur in constructors.

```
class Overload {
int  a;
void test(int x) {
a=x;
System.out.println("a: " + a);
}
void test(int x , int y) {
a= x;
int b= y;
System.out.println("a and b: " + a + "," + b);
}
```

}

```
class MethodOverloading {
public static void main( String args[ ] ) {
Overload overload = new Overload( );
overload. test(10);
overload. test(10, 20);

}
}
```

instance variable a is not mutually related to test( int x) or test( int x, int y ) as both are accessing the variable.

{ a } $\overline{R}$ { test( int x), test( int x, int y ) } --------------- 1

{ b } R { test( int x, int y ) }-------------------------------2

So, where 2 relations are found and we can say that these two relations will create two concepts.

## 3.2 Overriding

Overriding is runtime polymorphism. The methods are same in syntax. It is decided in the run time which method is to be invoked.

```
Method overriding.
class A {
int  i, j;
A( int a, int b ) {
i = a;
 j = b;
}

// display i and j
void show( )
{
System.out.println("i and j: " + i + " " + j );
}
}

class B {
int k;
B( int c) {
k = c;
}

void show( ){ // display k – this overrides show( ) in A

System.out.println("k: " + k);
  }
}

class Override {
public static void main(String args[ ] ) {
B subOb = new B(1 );
subOb.show( ); // this calls show( ) in B
}
}
```

A( ) accessing the instance variables i , j

A .Show( ) [ show( ) of class A ] accessing the instance variables directly for both A( ) and A. show( ) the relation comes like :

{i, j} $\overline{R}$ { A( ), A. show( ) } --------------------- 1

Similarly B( ) and B. show( ) exclusively related to variable k. so we can say that the relations are like this:

{B( ), B. Show( )} $\overline{R}$ {k} ----------------------- 2

So , two relations are creating two concepts. Now as the method show( ) is overridden, we shall consider the A. show( ) and B. show ( ) as a single entity say show( ). As we are considering here only the property of overriding we shall ignore the other methods and we can reduce the relations or dependencies like :

show( ) $\overline{R}$ { i, j, k } and therefore creating a module.

## 3.3 Inheritance

Inheritance is a property in Java where the members of a class inherit properties or attributes from its base classes. Inheritance can be of different forms multiple, hierarchical, multistage and hybrid.

```
class A {
    int x;
    int y;
    void showxy ( )
{
System.out.println(" x and y :  " + x + " " + y );
   }
   }
class B extends A
{
int z;
 void showz( ){
  System.out.println( "z : " + z );
   }
}

class Inheritance{
public static void main(String args[ ] )
{
  A a = new A( );
B b= new B( );
  a . x = 5;      // x of superclass A

  a.y = 5;        // y of superclass A

showxy ( );  // showxy ( ) of class A i.e. the superclass
   x= 10;     // x of subclass B as extended from A
   y= 10;    // y of subclass B as extended from A
   k= 10;    // k of own subclass B
 b.showxy( ); // showxy of superclass A extended by
           subclass B
   b. showz( );  // own method showz( ) of class B
     }
   }
```

x, y is mutually exclusively related to {showxy ( )} in case of class A and in case of class B too because showxy ( ) is inherited by class B from class A. So we can say that :

{x, y } $\overline{R}$ { showxy( ) }. Here one concept is created.

---------------------------- 1

Next, showz ( ) is accessing mutually exclusively to z. So, the relation goes like this :

{z } $\overline{R}$ {showz( )}. Another concept is created.--------- 2

As all relations are mutually exclusive we can take aggregation and can be written as:

{x, y, z }$\overline{R}${ showxy( ), showz( ) }----------------------- 3

## 3.4 Exception Handling

Exception handling is the property of java by which it can invoke some work when some normal task is prevented to execute by some faulty codes.

```
Class MyException {
    public static void main ( String args [ ] ) {
```

```
        int d, a;

        try {

        d= 0;

        a= 42 / d;

        System. Out. Println ( " This will not be printed. ");

        } catch (ArithmeticException e ) {  // catching of divided by zero errors

        System.out.println( " This is Division by zero creating an exception !!!" );

        }  System.out.println( " This is after the catch done…." );

                }

        }
```

In the class MyException a and d are instance variables. We can say that, try-catch block is directly accessing the variables because whenever the try block is executed then only the ArithmeticException e arises i.e. the instance variables are mutually exclusive with exception e. Thus they are creating concepts and therefore a module. By notation we write that :

$\{ a, d \} \overline{R} \{ \text{try-catch} ( ) \}$

### 3.5 Abstraction

Abstraction is the property of Java to hide details from the users so that the user can deal only with the functionalities of the codes.

```
 class Poly
 {
 // implementations and private members hidden

 Poly (int , int );
 double eval ( double );
 void add ( Poly );
 void mult ( Poly );
 public String toString ( );
 }

 public class Binomial
 {
 public static void main ( String[ ] args )

 {

 int N = Integer.parseInt ( args [ 0 ] ) ;

 double p = Double. parseDouble(args [1]) ;

 Poly y = new Poly ( 1, 0);
 Poly t = new Poly ( 1, 0);
 t.add ( new Poly ( 1, 1) );
 for ( int i = 0; i < N; i++ )
 {
 y. mult ( t ) ;
 Out.println(y + "");
 }
 Out.println("value: " + y.eval ( p ) );
 }
 }
```

Method add ( ) is using poly ( ) constructor. So, by notation we can write that :

$\{ \text{add} ( ) \} \overline{R} * \{ \text{poly} ( ) \}$ ---------------------- (1)

Moreover, the method add( ) and mult ( ) using the total class methods directly or indirectly as we can see from the class definition .Then also we can write that,

$\{ \text{add} ( ), \text{mult} ( ) \} \overline{R} * \{ \text{class Poly} \}$ ------------- (2)

It means that those methods are using the class methods otherwise.

## 4. Conclusion

We have tried to implement the basic properties of object oriented paradigm through concept analysis notation to create concepts as well as the modules. These modules will help us for reengineering because reengineering deals with change in the modules of codes to make the obsoluted or about to obsoluting software rework.

## Acknowledgment

All complements to the almighty who gives us the ability and knowledge to do this hard work. We express our greatest gratitude, appreciation and profound respect to our supervisor G Geetha, Dept. of Computer Applications, Lovely Professional University, Jalandhar, Punjab (India) for his continuous, proper & perfect guidance and invaluable suggestion.

## Authors' Profile

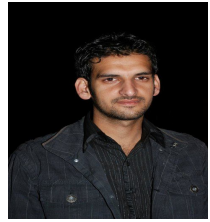

**Gulshan Kumar** pursuing his M. Tech degree in Computer Science and Engineering from Lovely Professional University, Jalandhar, India. His research interest includes Cryptography and Mobile Adhoc Networks.

**Mritunjay Kumar Rai** received his Ph.D. Degree from ABV-Indian Institute of Information Technology and Management, Gwalior, India. Currently he is working as an Assistant Professor in Lovely Professional University. His research interest area is Mobile Adhoc Networks and Wireless Sensor Networks.